# Why Do Weak-Binding M-N-C Single-Atom Catalysts Possess Anomalously High Oxygen Reduction Activity?

Di Zhang[1,3]*, Fangxin She[2], Jiaxiang Chen[2], Li Wei[2,]*, and Hao Li[1,]*

[1] Advanced Institute for Materials Research (WPI-AIMR), Tohoku University, Sendai 980-8577, Japan

[2] School of Chemical and Biomolecular Engineering, The University of Sydney, Sydney, New South Wales 2006, Australia

[3] State Key Laboratory of Mechanical System and Vibration, Shanghai Jiao Tong University, Shanghai 200240, China

*** Corresponding Authors**

Email: di.zhang.a8@tohoku.ac.jp (D. Z.)

Email: l.wei@sydney.edu.au (L. W.)

Email: li.hao.b8@tohoku.ac.jp (H. L.)

## Abstract

Single-atom catalysts (SACs) with metal-nitrogen-carbon (M-N-C) structures are widely recognized as promising candidates in oxygen reduction reactions (ORR). According to the classical *Sabatier* principle, optimal 3*d* metal catalysts, such as Fe/Co-N-C, achieve superior catalytic performance due to the moderate binding strength. However, the substantial ORR activity demonstrated by weakly binding M-N-C catalysts such as Ni/Cu-N-C challenges current understandings, emphasizing the need to explore new underlying mechanisms. In this work, we integrated a pH-field coupled microkinetic model with detailed experimental electron state analyses to verify a novel key step in the ORR reaction pathway of weak-binding SACs—the oxygen adsorption at the metal-nitrogen bridge site. This step significantly altered the adsorption scaling relations, electric field responses, and solvation effects, further impacting the key kinetic reaction barrier from HOO* to O* and pH-dependent performance. Synchrotron spectra analysis further provides evidence for the new weak-binding M-N-C model, showing an increase in electron density on the anti-bonding π orbitals of N atoms in weak-binding M-N-C catalysts and confirming the presence of N-O bonds. These findings redefine the understanding of weak-binding M-N-C catalyst behavior, opening up new perspectives for their application in clean energy.

## INTRODUCTION

The oxygen reduction reaction (ORR) is a crucial yet kinetically limited process in new energy technologies. The reliance on platinum-group metal (PGM) catalysts imposes substantial cost barriers that hinder widespread commercial implementation.[1, 2] Developing efficient, cost-effective non-PGM catalysts is essential to advance clean energy solutions,[3] especially for proton exchange membrane fuel cells (PEMFCs) and high-energy-density metal-air batteries, both of which are leading candidates for sustainable power sources in low-carbon clean transportation.[4, 5]

Single-atom catalysts (SACs) with metal-nitrogen-carbon (M-N-C) structures, where M typically represents 3*d* transition metals such as Mn, Fe, Co, Ni, Cu, and Zn, have demonstrated significant promise for enhancing catalytic performance in ORR.[6-8] Typically, the adsorption strength of ORR intermediates on 3*d* metal top sites diminishes as the number of *d*-orbital electrons increases. Consequently, M-N-C catalysts can be classified into three categories based on the adsorption strength of the metal sites toward adsorbates: strong-binding catalysts (M-N-C, with M=Ti, V, or Cr), moderate-binding catalysts (M=Mn, Fe, or Co), and weak-binding catalysts (M=Ni, Cu, or Zn). To highlight the experimental trends, we performed extensive data mining and gathered 1,018 experimental data points on M-N-C catalysts. (**Fig. 1**, with all source data stored in the Digital Catalysis Platform at https://www.digcat.org). It reveals that moderate-binding catalysts such as Fe/Co-N-C exhibit exceptional ORR activity in PEMFC applications, surpassing even that of PGM Pt/C catalysts.[9] This superior performance can be readily understood through the classic *Sabatier* principle and volcano models, which suggest that catalysts with moderate binding affinity are expected to deliver the best performance. However, weak-binding M-N-C catalysts, such as Ni-N-C,[10, 11] Cu-N-C,[12] and Zn-N-C,[13] demonstrate significant activity in alkaline media (**Fig. 1a** and **Fig. 1c**) and pH-dependent performance (**Fig. 1b** and **Fig. 1d**). These findings challenge current knowledge regarding M-N-C catalysts, as a comprehensive understanding of the unexpectedly high ORR activity and pH-dependent performance exhibited by weak-binding M-N-C catalysts remains unresolved.

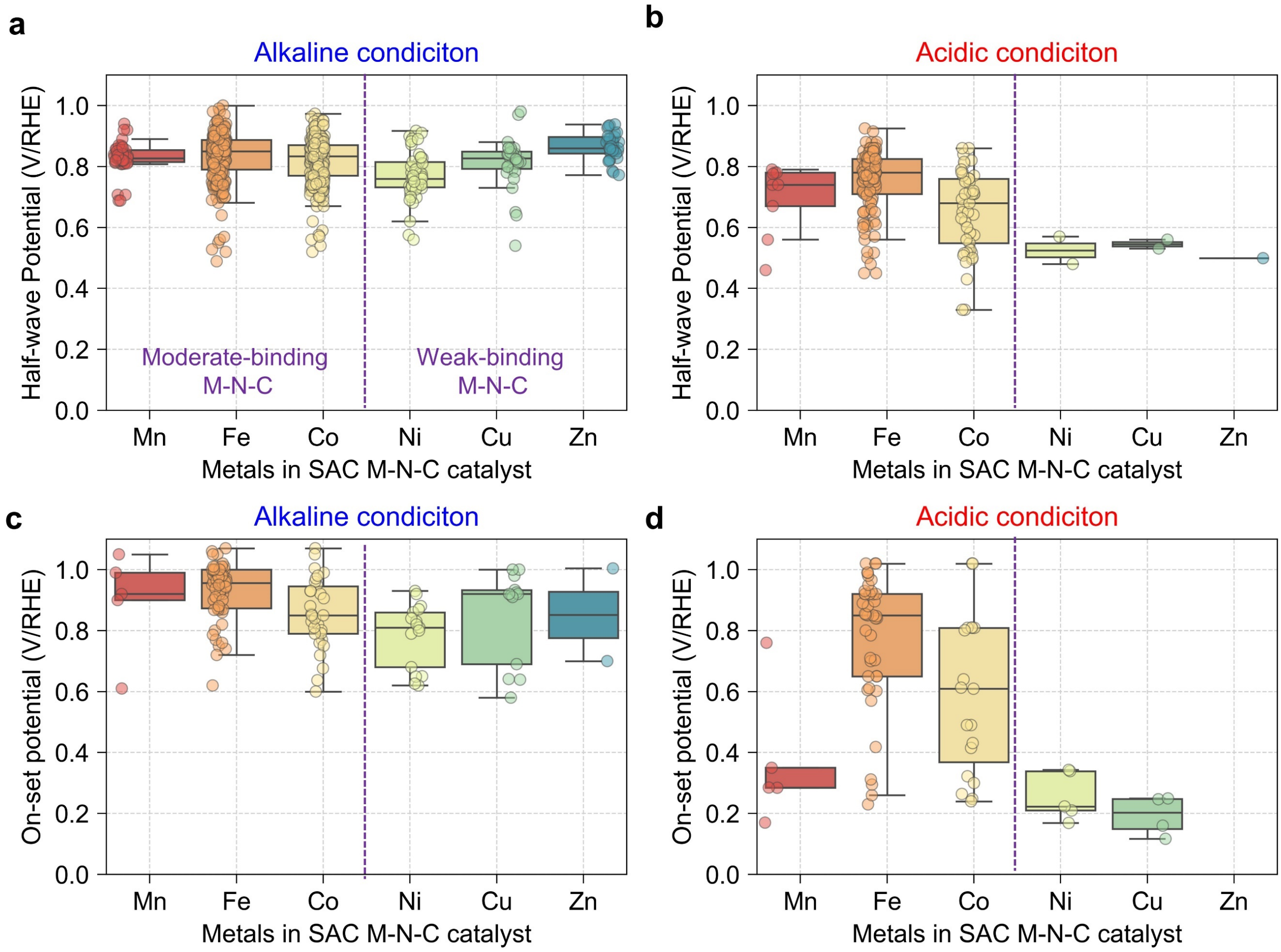

**Fig. 1. Summary of the ORR performance of M-N-C catalysts. (a-d)** represent the half-wave potential under (**a**) alkaline and (**b**) acidic conditions, and the on-set potential under (**c-d**) both conditions (on-set current density is generally reported as 0.5 mA $cm^{-2}$). Big data analysis indicates that weak-binding M-N-C catalysts (Ni/Cu/Zn-N-C) exhibit excellent ORR performance under alkaline conditions and show pronounced pH dependence (The above data are stored in https://www.digcat.org).

To understand the intrinsic activity of weak-binding M-N-C catalysts, it is crucial to first eliminate the influence of structural defects in SACs. The metal atop site of SACs is typically considered as the active center for ORR.[14] However, some studies have suggested that surrounding nitrogen or carbon atoms can also serve as active sites, especially when M-N-C catalysts contain nanopore defects.[15, 16] To mitigate the impact of defects or uncertain structures, recent studies[11, 17-19] have often used heterogenous molecular M-N-C catalysts, which are obtained by depositing organometallic molecules, such as metal porphyrins (MPOR) and phthalocyanines (MPc) on a carbon substrate or incorporating them into metal-organic frameworks (MOFs) and covalent organic frameworks (COFs). These catalysts are preferred due to their well-defined atomic structure and the improved dispersion of active sites.[17-20] Therefore, using molecular M-N-C catalysts instead of those prepared by direct high-temperature pyrolysis helps theoretical modeling to identify the true sources of reaction activity, including the active sites or reaction pathways.

In this work, we combined a universal pH-field coupled microkinetic model with detailed experimental electron state analyses to confirm the critical step in the reaction pathway of weak-binding systems—the stabilization of single-atom oxygen at the metal-N bridge site. For weak-binding M-N-C SACs, we discover that the atomic oxygen (O*) in ORR tends to spontaneously adsorb at the M-N bridge-sites during ORR. The bridge-adsorbed O* exhibits unique features that lead to significant discrepancy to the moderate-binding M-N-Cs (*e.g.*, M = Fe or Co), including different adsorption scaling relations, distinctive response to the electric fields, and different solvation effects. Furthermore, kinetic analysis revealed that the O* adsorbed at the bridge-sites can significantly lower the kinetic barrier of HO-O* bond activation during ORR. By including these missing mechanisms, a pH-dependent microkinetic volcano specifically for weak-binding M-N-C SACs was established, showing good agreements with experimental observations. Our model significantly enhances the accuracy in predicting the activity of weak-binding M-N-Cs in both acidic and alkaline media. Synchrotron spectral analysis provides strong evidence for our weak-binding M-N-C model, revealing increased electron density in the anti-bonding $\pi$ orbitals of N atoms in Cu/Ni-N-C after ORR testing, confirming the formation of N-O bonds. This critical mechanism will be instrumental in facilitating the better refined and targeted design of M-N-C catalysts in future research endeavors.

## Methods

**Computational methods.** DFT calculations for ORR adsorbate binding energies were conducted utilizing the generalized gradient approximation approach, specifically employing the revised

Perdew−Burke−Ernzerhof (RPBE) functional for the representation of electronic exchange and correlation effects.[21, 22] The RPBE functional addresses overbinding issues in traditional GGA functionals like PBE, making it well-suited for systems with weak to moderate adsorption.[23] For ORR and similar reactions, RPBE reliably estimates the adsorption energies of intermediates like O*, HO*, and HOO*, enabling accurate predictions of reaction energetics and pathways.[23] Core electrons were modeled using the projector augmented-wave method,[24] while valence electrons were represented through a plane-wave basis set expansion of Kohn–Sham wavefunctions,[25] applying a minimum cutoff energy of 400 eV. For 3d transition metal ions, our DFT calculations include spin polarization, allowing for the independent optimization of spin-up and spin-down electron densities. The initial spin state of the 3d metal ion is set based on its prior experimental or theoretical knowledge. External electric fields, ranging from −0.6 to 1.0 V/Å, were systematically applied. For each field setting, adsorbate structures underwent relaxation until achieving a force convergence criterion of 0.05 eV/Å. The most stable conformations under each field condition were selected to estimate adsorbate energies. Detailed descriptions of the computational approaches and modeling techniques are available in the **Supplementary Methods** section. The COHP analysis in this study was conducted using the LOBSTER toolkit. [26] The climbing image nudged elastic band (CINEB) method,[27] implemented in the ASE framework, was used to identify the transition state along the reaction pathway. A total of 5 images, including the initial and final states, were used to sample the pathway. The calculations were performed using the force and energy convergence criteria consistent with DFT calculations.

**Synthesis of M-COF366/CNT and MPc/CNT catalysts.** All chemicals were purchased from Sigma-Aldrich (otherwise stated) and used without further treatment. MWCNTs were obtained from CNano. The COF-366 decorated MWCNT catalysts were synthesized using a solvothermal approach. Initially, 0.02 mmol of 5,10,15,20-(tetra-4-aminophenyl) porphine (TAPP, Porphychem, >98%) and 0.04 mmol of terephthaldehyde (TPD, 99%) were dissolved in a mixture containing 1 mL of ethanol (absolute, 200 proof), 1 mL of mesitylene (98%), and 0.1 mL of 6 M acetic acid (99%) inside a Pyrex tube. To this, purified carbon nanotube was added in a 2:1 mass ratio relative to the COF-366 precursors. This mixture was then uniformly dispersed via 30 minutes of sonication. Subsequently, the suspension underwent a series of three freeze-thaw cycles using liquid nitrogen, followed by a preheating step at 65 °C for 4 hours under an argon atmosphere before it was sealed and solvothermal treated at 120 °C for 72 hours. Afterwards, the solid product was isolated through filtration, thoroughly washed with ethanol and N, N-dimethylacetamide (DMAC, ≥99%), and then vacuum-dried. This product, COF366/CNT, was utilized as the base for the M-COF366/CNT catalysts synthesis. Multiple iterations of this synthesis procedure were conducted to accumulate sufficient material for the subsequent metalation stage.

Metalation of COF366/CNT involved using various metal acetate salts ($M(Ac)_2$, M=Ni and Cu, 99.9 metal traces). Approximately 45 mg of the pre-prepared COF366/CNT and 0.27 mmol of metal salts were dispersed in 5 mL of methanol, followed by the addition of 20 mL of chloroform (≥99%) and 15 mL of dimethylformamide (DMF, ≥99.9%). The mixture was subjected to an hour of bath sonication and then continuously stirred at 80 °C for 24 hours under argon. After cooling, the product was filtered, washed with deionized water, and dried in a vacuum. The resultant catalysts are denoted as M-COF366/CNT, with M=Ni or Cu.

The MPc/CNT catalysts were prepared by loading various MPc (M = Ni, Cu, Porphychem, ≥95%) onto a purified MWCNT using a method previously reported in our study.[18] In this method, 3 mg of MPc and 20 mg of purified MWCNT were combined in 20 mL of DMF, followed by 30 minutes of ultrasonication. The mixture was then stirred for 24 hours under an argon atmosphere. The solid product was later isolated, washed with DMF and ethanol, and dried under vacuum at 80 °C overnight, resulting in the MPc/CNT catalysts.

**Characterization and Electrochemical Methods.** To assess metal residue in purified CNT and metal content in various catalysts, inductively coupled plasma atomic emission spectroscopy (ICP-AES) was performed on a Perkin Elmer Avio 500 spectrometer. Samples were acid-digested in 6 M $HNO_3$, followed by appropriate dilution.

Transmission electron microscopy (TEM) and energy-dispersive X-ray (EDX) analyses were conducted using an FEI Themis-Z microscope, in both bright-field high-resolution (BF-HRTEM) and high-angle annular dark-field scanning (HAADF-STEM) modes. X-ray photoelectron spectra (XPS) were collected on a Thermo Scientific K-Alpha+ spectrometer with an Al K$\alpha$ source (1486.3 eV). Spectra energy was calibrated with the C 1*s* line collected from a graphite standard. A pass energy of 20 eV is used. Spectra were analyzed using a using Shirley-type background. Detailed peak fitting parameters are listed in the Supporting Information (**Table S5**). Synchrotron-based X-ray absorption spectroscopy (XAS) data were collected at the ANSTO XAS Beamline of the Australian Synchrotron. The electron beam was generated using a liquid nitrogen-cooled Si(111) monochromator with Si-coated collimating and Rh-coated focusing mirrors, producing a beam size of approximately 1 × 1 $mm^2$. Measurements were taken in transmission mode, with energy calibration performed using metal foils. Spectral processing and fitting were conducted in the Demeter package with the FEFF 9.0 code.

The electrochemical performance of the catalysts was evaluated using an Autolab PGSTAT302 electrochemical workstation and Pine Research MSR rotator in a three-electrode system at 25 °C. The setup included a rotary ring-disk electrode (RRDE, E6R1PK, Pine Research) with a glassy carbon disk (5 mm diameter) and a Pt ring. The collection efficiency was calibrated by $Fe(CN)_6^{-3/-4}$ redox reaction and found 0.249. The electrode was polished using 1.0, 0.03, and 0.05 μm $Al_2O_3$ powder prior to each test. Reference electrodes used were a pre-calibrated Ag/AgCl (3M KCl) electrode for acidic conditions and a Hg/HgO (0.1 M KOH) electrode for alkaline conditions. The counter electrode was a graphite rod (AFCTR3B, Pine Research). The electrode was set at a speed of 1600 rpm. Linear sweep voltammetry (LSV) curves were collected at a scan rate of 5 mV $s^{-1}$ without iR-compensation. The on-site potential was determined from the electrode potential corresponding to a kinetic current density of 0.1 mA $cm^{-2}$. All potentials were normalized to the reversible hydrogen electrode ($V_{RHE}$). Details on the calculation methods for kinetic current densities and TOFs are provided in the **Supplementary Methods**.

## Results and Discussion

### Bridge adsorption and activity descriptor

We performed DFT calculations on over 50 M-N-C (where M = Fe, Co, Ni, and Cu) structures, including pyrrolic and pyridine (**Fig. 2a**) and molecular M-N-C catalysts (**Fig. 2b**). ORR adsorbate configurations including HO*, O*, HOO*, $O_2$*, and $H_2O_2$* were optimized on these M-N-C catalysts. To identify the preferred adsorption site, we initially positioned ORR adsorbates either on the metal atop site or the M-N/C bridge site, starting at a distance greater than 2.5 Å from the substrate. For Fe-N4 structures, even when the adsorbates were initially placed above the bridge sites, they spontaneously shifted to the metal atop site. In contrast, some ORR adsorbates, such as O* and HO* on Cu/Ni-N-C, remained at the bridge site when placed there initially. As shown in Supporting Information **Fig. S1**, we performed comparative climbing-image nudged elastic band (CI-NEB) calculations under both solvent and solvent-free conditions for the reaction pathway from atop O* to bridge O*. For Cu-pyrrole-N, the energy barrier for this transition is negligible under both conditions, suggesting the spontaneous transfer of oxygen atoms to bridge-site adsorption on Cu-pyrrole-N. In the case of Ni-pyrrole-N, a small energy barrier of approximately 0.2 eV is observed for the transition from atop O* to bridge O*. Furthermore, we found that this barrier remains nearly unchanged under both solvent and solvent-free conditions. To clarify whether the atop- or bridge-site is more thermodynamically stable for HO* and O*, **Fig. 2c** illustrates the more favorable site for O* and HO*. The arrow indicates that O* and HO* are more likely to adsorb on the M-N/C bridge-site when there are fewer nitrogen atoms around or when the central metal atoms do not bind them tightly, such as metal atoms Ni and Cu. Meanwhile, pyridinic (pyd in **Fig. 2c**) and pyrrolic (pyr in **Fig. 2c**) M-N-C catalysts exhibit minimal differences in bridge adsorption. Among catalysts with the same metal atoms and nitrogen coordination, weak-binding pyrrolic M-N-C appears more likely to adsorb HO* or O* at the bridge site. Particularly for the M-N4 structures, O* prefers to adsorb on the bridge site of pyrrolic structures but on the top side of pyridinic structures. Therefore, we primarily employ pyrrolic-type M-N-C structures in the following electric field simulations and experimental tests.

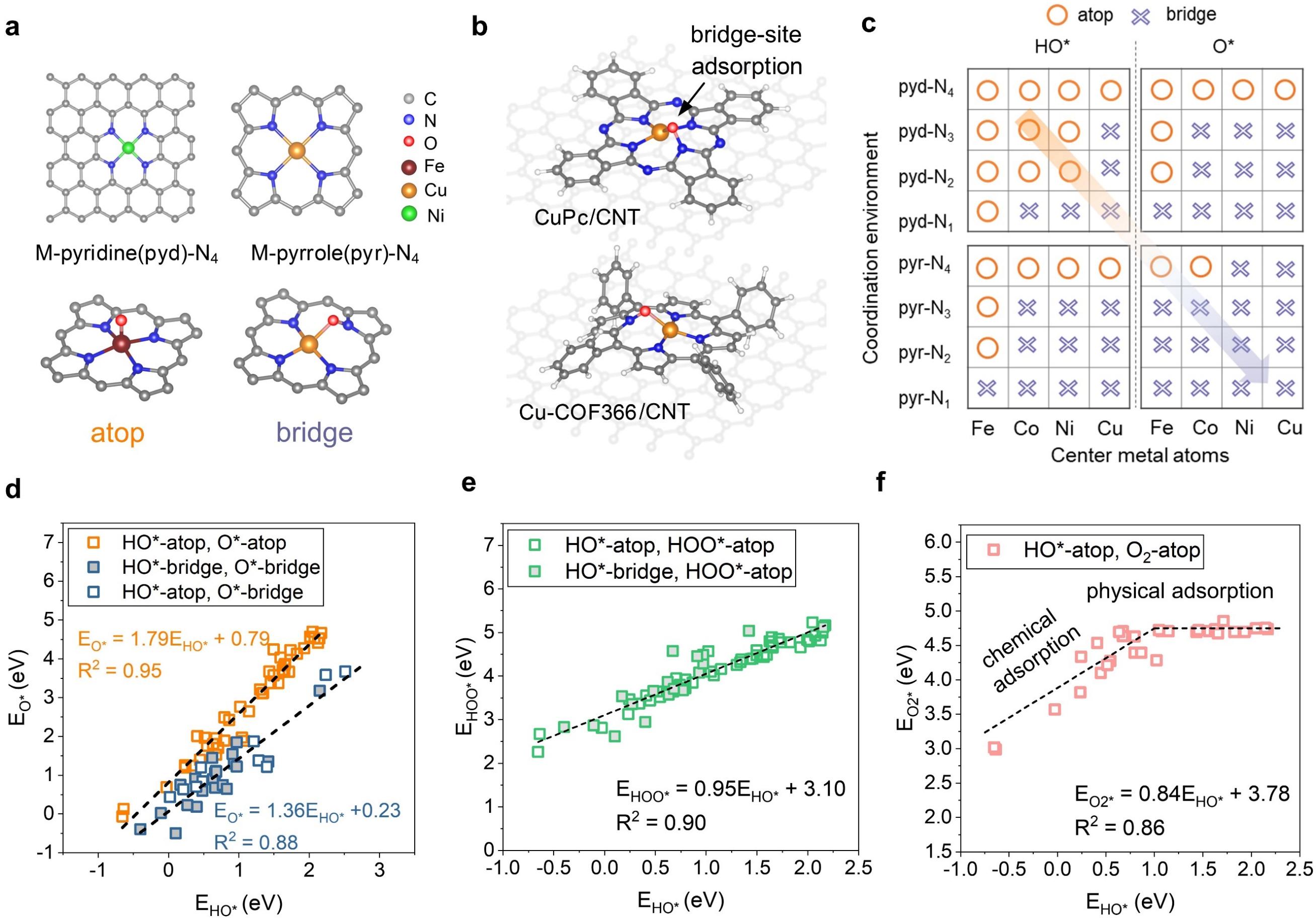

**Fig. 2. Analysis of the favorable adsorption sites and linear scaling relations that determine ORR activities.** (**a**) Atop- and bridge-site adsorption on M-pyridine-N and M-pyrrole-N structures. (**b**) The identified favorable bridge-site O* adsorption on Cu-phthalocyanine (CuPc) and Cu-covalent organic framework-366 (Cu-COF366). (**c**) Variation in the most favorable adsorption sites with different central metals (Fe, Co, Ni, and Cu) and nitrogen coordination numbers (pyd-$N_{1\sim4}$ and pyr-$N_{1\sim4}$). Scaling relations between (**d**) $E_{O*}$ and $E_{HO*}$, (**e**) $E_{HOO*}$ and $E_{HO*}$, and (**f**) $E_{O2*}$ and $E_{HO*}$. (The $E_{H2O2*}$-$E_{HO*}$ scaling relation is detailed in **Fig. S2**.)

Since the most favorable sites for ORR adsorbates on M-N-C catalysts are not consistently the metal atop-site, we categorized the data into three groups, as labeled in **Fig. 2d**. The adsorption energy of O* *versus* HO* fits into two distinct scaling relations. In weak-binding systems, the adsorption energy of oxygen at bridge sites can be significantly lower (more than 1.0 eV) compared to that at atop sites, which could substantially alter the ORR activity. Regarding the adsorption energy between HOO* and HO* (**Fig. 2e**), the data largely obey to the universal HO* *versus* HOO* scaling.[28] As for the adsorption energies of $O_2$* ($E_{O2*}$, **Fig. 2f**), a clear scaling relationship is evident when the adsorption is relatively strong, corresponding to chemical adsorption. In contrast, on surfaces with weak adsorption, $E_{O2*}$ tends to stabilize at around 4.75 eV (the horizontal dashed line in **Fig. 2f**), corresponding to physical adsorption. We also established a correlation between the adsorption energy of $H_2O_2$* and that of HO*, resulting in a quasi-horizontal linear relation: $E_{HOOH*} = 0.03E_{HO*} + 3.45$ (**Fig. S2**). When considering bridge-site adsorption, the most significant variation among the four linear scaling relations occurs in the adsorption energy of the atomic O*, emphasizing the importance of the bridge-adsorbed O* in the ORR process on these weak-adsorption systems.

**Modelling pH with electric fields**

In traditional pH-dependent models, researchers typically simulate pH by adjusting the work function (corresponding to $U_{SHE}$) of the catalyst surface under a given $U_{RHE}$ condition, generally by adding alkali metal ions or electrons.[29, 30] However, recent studies have shown that modulating the electric field provides a faster and accurate way to describe the pH dependence in electrocatalytic reactions.[19, 31] This approach forms the basis of the pH-electric field coupled microkinetic model[32] (see **Supplementary Methods** Section S1 for more details). So first, we calculated the changes in adsorption energy across an electric field ranging from −0.6 to 1.0 V Å$^{-1}$. This range is determined by the potential of zero-charge (PZC) of M-N-C catalysts (**Fig. S3**). **Fig. 3a** illustrates the response of atop-site-adsorbed ORR adsorbates to varying electric fields. Notably, O* and HOO* exhibit a significantly stronger response on Cu/Ni-pyrrole-N surfaces. During the ORR process, the thermodynamic driving force for breaking the O-O bond in HOO* primarily depends on the free energy change of the reaction step HOO* + $H^+$ + $e^-$ →O* + $H_2O$. Theoretically, a lower $G_{HOO*}$ (indicating stronger adsorption) or a higher $G_{O*}$ (indicating weaker adsorption) reduces the overall change in free energy, thereby weakening the thermodynamic driving force. This explains why, under acidic conditions, the transition from HOO* to O* becomes the rate-limiting step in many M-N-C ORR catalysts.[33] Interestingly, as shown in **Fig. 3b**, when O* is adsorbed on the M-N/C bridge site, the change in adsorption free energy is minor. This may enhance the thermodynamic driving force for the breaking of the O-O bond in HOO*, leading to a catalytic activity distinctly different from that at the atop-site.

To understand the distinct response of adsorbed O* to the electric field at the atop- and bridge-sites, we also analyzed the charge density difference resulting from the adsorption of O*. We observed that the dipole moment change induced by the atop-adsorbed-O* is perpendicular to the graphene surface, aligning with the direction of the electric field (**Fig. 3c**). In contrast, the dipole moment changes upon bridge O* forms an angle with the electric field, weakening the electric field's impact on O* adsorption energy. According to our pH-field dependent model, the pH dependence is primarily attributed to the different responses of adsorption energies of ORR intermediates to the electric field. When the adsorption energy of O* is less sensitive to the electric field, the reaction rate of the rate-determining step from HOO* to O* will also exhibit smaller variations with the electric field, and consequently, with pH. Therefore, our findings suggest that changing the orientation of the dipole moment in ORR adsorbates, especially for O*, could reduce the pH dependence of the catalyst.

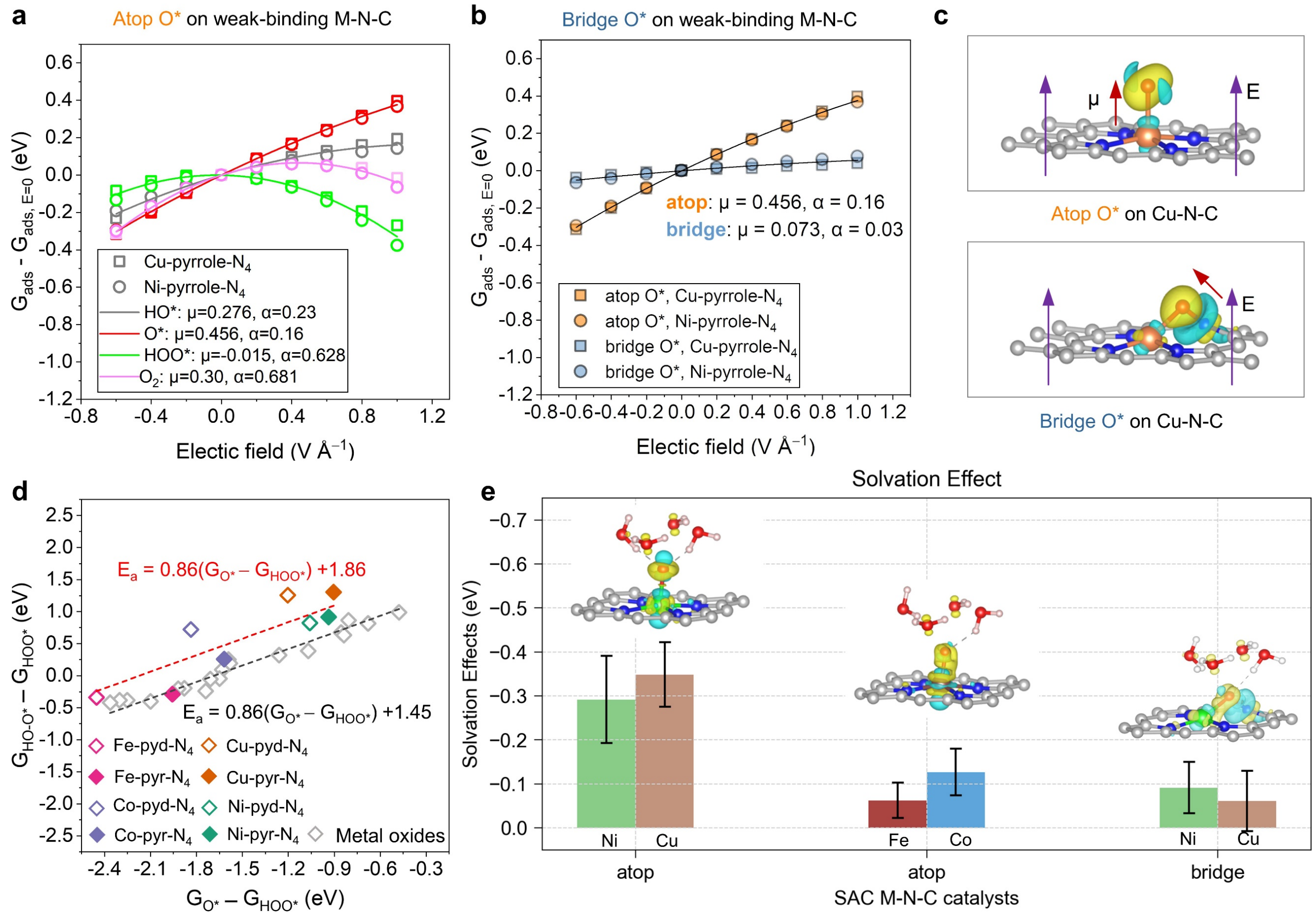

**Fig. 3. pH-electric field coupled microkinetic modeling for weak-binding M-N-C catalysts.** (**a-b**) Field effects on (**a**) atop and (**b**) bridge ORR adsorbates with fitted values for dipole moment μ (e) and polarizability α ($e^2$ $V^{-1}$) for Ni-pyrrole-$N_4$ and Cu-pyrrole-$N_4$. (**c**) Charge density difference induced by the adsorption of atomic O* on the atop- and bridge-sites, respectively. Yellow and teal colors in the isosurfaces represent electron charge gain and loss, respectively. The red and purple arrows represent the dipole moment (μ) and electric field (E), respectively. The dipole arrow follows the convention in chemistry, representing the dipole moment goes from positive to negative center. (**d**) Calculated kinetic energy barriers of HO-O* activation on M-N-C SACs and the linear relations between the transition state energy of HOO* activation and the free energy difference of $G_{O^*}$ and $G_{HOO^*}$. (**e**) Solvation effects of atop-adsorbed O* on Fe/Co/Ni/Cu-pyrrole-$N_4$ and bridge-adsorbed O* on Ni/Cu-pyrrole-$N_4$. The electron transfer between the metal and single-atom oxygen. (O*) substantially impacts the solvation effect of oxygen.

The adsorption of O* at the bridge sites on the surfaces of weak-binding M-N-C SACs also significantly influences the kinetic barriers during ORR. The intrinsic barriers for all proton-electron transfer steps were set at 0.26 eV and assumed to be field-independent, aligning with the findings of Tripković *et al*.[34] and Hansen *et al*.[35] However, it is suggested that the activation energy to start the protonation of HOO* to O* could be higher due to the breaking of O-O bond.[36] To construct a unified kinetic model, we calculated the transition state (TS) energies on M-pyridine/pyrrole-N (M = Fe, Co, Ni, and Cu) using the CI-NEB method[27] using explicit solvation models. Since atomic oxygen is more stable at the bridge-site of weak-binding SACs, we looked at two types of final states (FS), which are based on where the O* is adsorbed - either at the atop- or the bridge-site. **Fig. S4** shows the activation barriers for the HO-O bond breaking at different catalysts. Fe-pyrrole/pyridine-N typically has a strong ability to activate the HO-O* bond, resulting in an almost negligible activation energy barrier.[37] While on the atop-sites of weak-binding surfaces like Cu/Ni-N-C, there is an activation energy barrier >1.0 eV, suggesting that the activity at the atop-site of Cu/Ni-N-C might be very low. However, if the movement of O* from the atop-site to the bridge-site is considered, the energy barrier for HO-O* activation would not exist (see **Fig. S4**). Next, to construct an activity model using a single descriptor, we fitted the TS energy as a linear function of the difference between $G_{O^*}$ and $G_{HOO^*}$ (**Fig. 3d**), widely known as the Brønsted-Evans-Polanyi (BEP) relation.[38] In the meantime, we compared the BEP relation of M-N-C catalysts against that derived from transition metal oxides in the reference.[33] These data points predominantly adhere to a consistent linear scaling relation. Yet, the activation energy for HOO* on M-N-C catalysts is about 0.4 eV higher than that observed on metal oxides. Consequently, the transformation process from HOO* to O* is inferred to be considerably sluggish on the atop sites of weak-binding M-N-C catalysts. This finding further substantiates the inference that the adsorption of O* on the bridge-sites of weak-binding M-N-C SACs significantly influences the reaction dynamics, both kinetically and thermodynamically. Some previous studies have also reported the phenomenon of co-adsorption at metal sites in single-atom M-N-C materials.[14] Therefore, we constructed models for ORR activity at metal sites under conditions of top and bottom bridge-site oxygen adsorption (**Fig. S5**). However, we found that when both the thermodynamic free energies and reaction barriers along the reaction pathway are taken into account, reaction pathways under co-adsorption conditions typically exhibit lower reaction activities due to the higher energy barriers. A more detailed discussion can be found in Supporting Information (**Fig. S5**).

Elucidating the solvation effects on electrochemical reaction energetics is crucial in understanding the electrocatalytic reactions.[39] The solvation correction is primarily derived from the hydrogen bonding between ORR adsorbates and the surrounding water molecules. Each hydrogen-bond in an aqueous solvent is expected to contribute approximately 0.15 eV of energy, based on an enthalpy difference of 0.45 eV when water transitions from the gas to the liquid phase.[40, 41] Ideally, this would lead to solvation energy corrections of 0.05, 0.35, and 0.40 eV for the O*, HO*, and HOO*, respectively. Yet, actual solvation effects are usually influenced by the charge states of the ORR adsorbates. To investigate the solvation effects on the adsorption energy of O*, as well as the adsorbates for HOO*, HO*, and $O_2$* (Source data is available in **Fig. S6** and **Table S2**), we examined the variation in solvation energies across different solvation conditions by introducing an increasing number of water molecules to the adsorbed intermediates. This approach is also typically used for calculating solvation energies due to its lower computational demand compared to *ab initio* molecular dynamics (AIMD) simulations.[42] **Fig. 3e** shows that the weak adsorption surface, like Cu and Ni-pyrrole-$N_4$, exhibits a significantly higher solvation energy, ranging from –0.3 to –0.4 eV. In contrast, the average solvation effects of O* on Fe-pyrrole-$N_4$ and Co-pyrrole-$N_4$ surfaces are merely –0.08 eV. Further charge density difference analyses suggest that this discrepancy may stem from the distinct charge transfer between Fe/Co-O and Ni/Cu-O bonds, as depicted in the inset of **Fig. 3e**. When atomic oxygen adsorbs on Fe/Co-pyrrole-$N_4$, the electrons obtained by O* are primarily distributed between the metal and O*. Conversely, for O* adsorbed on the weak-binding Cu/Ni-pyrrole-$N_4$, the increased electrons near O* are mainly localized around O*. This leads to a significant increase in the strength of H-bonding between the oxygen and the hydrogen atoms of water molecules in the solvent. However, if O* moves from the atop-site to the bridge-site, the electrons received by the O* will once again be primarily distributed between O* and the metal atom, as shown in right most of **Fig. 3e.** Therefore, the solvation effect of O* adsorbed at the bridge-site is significantly reduced compared to the solvation effect of the O* adsorbed at the metal atop-site. This accurate comprehension of the solvation mechanism on the weak-binding M-N-C surfaces is essential for the development of more precise theoretical models.

As illustrated in **Fig. 4**, the theoretical calculations presented above allow us to construct two distinct activity volcano models. These models incorporate different scaling relations, electric field responses, and solvation effects, each specifically tailored to the atop-site O* and bridge-site O* configurations, respectively. The activity volcanos exhibit a clear pH-dependent turnover frequency (TOF): the increase in the electric field (*i.e.*, shifting from alkaline to acidic conditions) leads to the decreasing catalytic activities. The electric field range of 0.5–1.5 V/Å shown in **Fig. 4** was calculated using Equation (6) in the Supporting Information. For pH values ranging from 0 to 14, the electric field on the surface of these materials under ORR electrode potentials (0.6 V) is approximately 0.5–1.5 V $Å^{-1}$, with CH=25 μF $cm^{-2}$ and $U_{PZC}$=−0.47 V/SHE. Additionally, the shape of the activity volcano is distinctly different for atop-O* and bridge-O* processes. In **Fig. 4a**, the activity volcanoes representing the 4e- ORR process (*i.e.*, from $O_2$ to $H_2O$). The 4-electron activity declines sharply in both alkaline and acidic solutions when $G_{HO*}$ exceeds 0.8 eV, as shown in **Fig. 4a.** This decline is attributed to the elevated HOO*-O* energy barrier at the atop-site (**Fig. S4**). Conversely, the bridge-site O* process, as shown in **Fig. 4b**, shows greatly enhanced activity around $G_{HO*}$ = 2.0 eV compared with the atop-site activity volcano (**Fig. 4a**). The significant difference in the activity volcano is primarily due to the stronger binding of O* on the Cu/Ni-N bridge site, which greatly lowers the energy barrier for O-O bond breaking in HOO*, as shown in **Fig. 3d**. Additionally, the single oxygen atom adsorbed on the bridge site exhibits a lower dipole moment and polarizability (**Fig. 3b**), further enhancing the reaction from HOO* to O*. In **Figs. 4c** and **4d**, we also derive the 2e- ORR activity volcanoes and observe that the peaks for the atop and bridge sites are located in different regions. For the atop site, the optimal range of $G_{HO*}$ is from 1.5 to 1.8 eV, while for the bridge site, the optimal range shifts to 2.0 to 2.5 eV due to the enhanced reaction from HOO* to bridge-O*. Based on the experimental observations (**Fig. 1**), the new weak-binding model provides a better explanation.

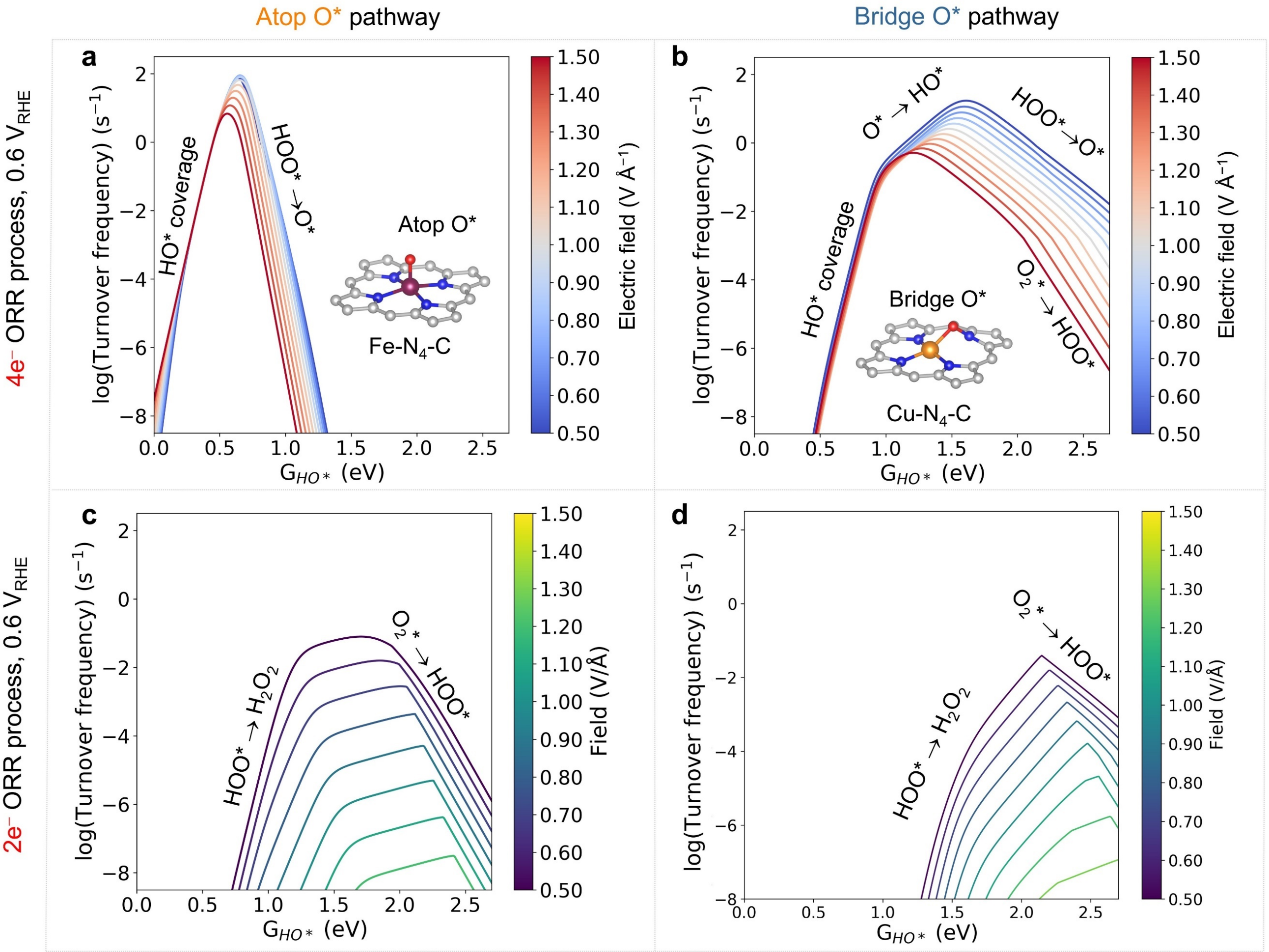

**Fig. 4. pH-Dependent activity volcanoes for the atop-O* and bridge-O* reaction mechanisms.** (**a-b**) 4e- ORR process *via* (**a**) atop-O* and (**b**) bridge-O* at 0.6 $V_{RHE}$. (**c-d**) 2e- ORR Process *via* (**c**) atop-O* and (**d**) bridge-O* at 0.6 $V_{RHE}$.

To validate the new reaction mechanisms, two different sets of catalysts were prepared by either depositing metal phthalocyanine (MPc, M=Cu and Ni) on purified carbon nanotubes (denoted as MPc/CNT) or a porphyrin-based covalent organic framework (COF-366) coated CNTs (denoted as M-COF366/CNT, see synthesis details in **Method**). High-angle annular dark-field scanning transmission electron microscopy images (HAADF-STEM) of these catalysts are collected. **Figs 5a-b** show the results of NiPc/CNT and Ni-COF366/CNT, respectively, and others are displayed in **Fig. S7**. These catalysts display clear carbon nanotube graphitic lattice fringes with high-contrast points sparsely distributed on it, indicated by red circles, which can be assigned to the metal atoms in the organometallic molecules. Despite the COF shell being sensitive to electron beams, bright spots are still visible, attributed to single metal atoms. X-ray energy dispersion elemental mapping also afforded uniform distribution of carbon (C), nitrogen (N), and various metal elements in these catalysts (**Fig. S8**). The synchrotron illuminated metal K-edge X-ray absorption near-edge structure (XANES) spectra, as shown in **Figs 5c-d,** and their extended X-ray absorption fine structure (EXAFS, **Fig. S9** and **Table S3**) fitting results confirmed that the metals are solely distributed in the M-$N_4$ configuration.

Next, we collected the ORR performance of these catalysts in acidic (0.1 M $HClO_4$, pH=1.3) and alkaline (0.1 M KOH, pH=12.6) electrolytes on a rotary ring-disk electrode with pre-calibrated collection efficiency. The ORR performance of the pristine CNT and non-metallated COF366/CNT substrates was also tested and used to correct the catalysts (**Fig. S10a**), since both substrates exhibit some ORR performance in alkaline electrolyte, although inferior to the catalysts (**Fig S10b**). We calculated the oxygen-diffusion corrected kinetic current density ($j_k$, **Fig. 5e**) using the Koutecky-Levich equation (see **Supplementary Methods, Section S3** for detailed calculation methods) for a meaningful comparison between experimental and theoretical results. Further in **Fig. 5f,** we present the simulated kinetic current densities for weak-binding SACs, with calculation methods detailed in the **Supplementary Methods, Section S2**. These simulations accurately reflect the experimental ORR activities shown in **Fig. 5e**, particularly under alkaline conditions (pH = 12.6). Additionally, the trend in activity across different pH levels observed in the simulations is consistent with the experiments.

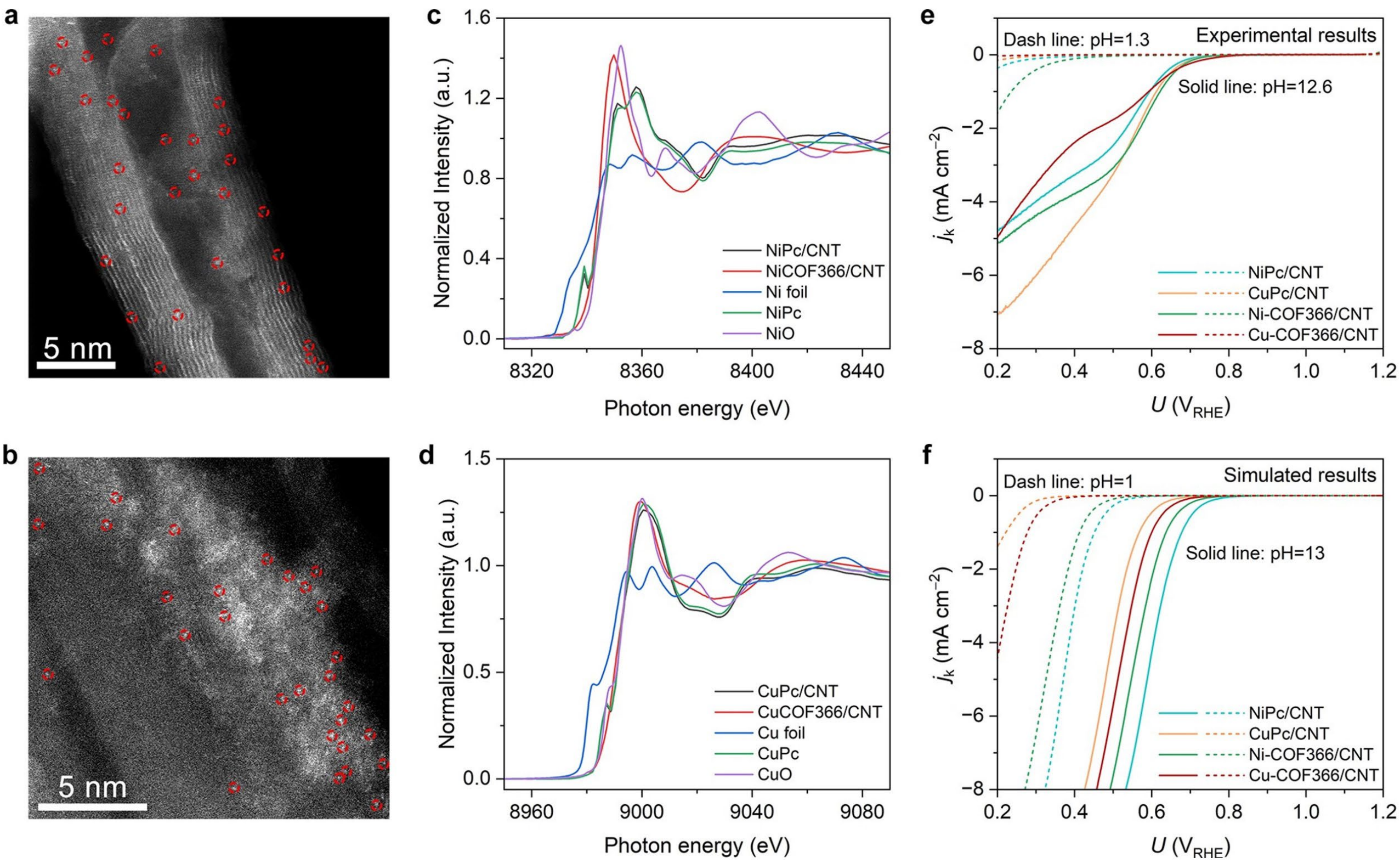

**Fig. 5. Catalyst characterizations and ORR performance of the weak-binding SACs.** The HAADF-STEM image of (**a**) NiPc/CNT; (**b**) Ni-COF366/CNT. The HAADF-STEM images of CuPc/CNT and Cu-COF366/CNT are presented in **Fig. S7**. The high-contrast points are sparsely distributed on the surface (marked by the red circles), which can be assigned to the metal atoms in the M-N-C catalysts. Metal K-edge X-ray absorption near-edge structure (XANES) spectra (**c**) Ni-based and (**d**) Cu-based samples and their references. (**e**) Experimental and (**f**) simulated $j_k$ at pH 1.3 (dashed line) and 12.6 (solid line) of the molecular M-N-C catalysts.

To further demonstrate the reliability of the new model**, Fig. 6a** presents a correlation plot of the onset potential obtained from both experiment and simulation, demonstrating a quantitative agreement between the experimental and theoretical values. For the experimental benchmarking of the activity volcanos, **Fig. 6b** presents a comparative analysis of two sets of pH-dependent kinetic catalytic activity volcano plots. These plots illustrate the 4-electron (4e-) pathway of M-N-C catalysts at 0.6 $V_{RHE}$ *via* atop O* and bridge O* reaction processes, respectively. The scatter points in **Fig. 6b** represent the TOF obtained experimentally. For the calculations of TOF, it is assumed that all the metals identified from ICP-AES (Table S4) analyses for the MPc/CNT catalysts were active for ORR since they are molecularly dispersed on the catalyst surface. The exposed metal sites in the M-COF366/CNT catalysts were further evaluated by a nitrite stripping method,[43, 44] which revealed the number of electrolyte-accessible metal active sites (**Fig. S11** and **Table S4**). It is evident that the experimental data align more closely with the kinetic model of O* adsorbed at the bridge site (**Fig. 6b**). Both in acidic (red dots) and alkaline (blue dots) environments, the results show qualitative and quantitative agreement with the theoretical curves. Therefore, this also demonstrates the reliability of the proposed ORR mechanism for weak-binding M-N-C catalysts and the efficacy of combining theoretical and experimental methods to determine reaction mechanisms. Notably, the Helmholtz capacitance ($C_H$) in our pH-dependent model introduces some uncertainties. To address this, we conducted additional uncertainty tests on $C_H$ using a theoretical framework presented in a previous study,[45] which proposed that the inverse of $C_H$ is proportional to the vacuum layer thickness at the solid-liquid interface. In our previous work, we performed AIMD simulations to systematically investigate the interface between M-N-C catalysts and water solvent.[46] These simulations suggested that the vacuum layer thickness at the solid-liquid interface for weakly adsorbing systems ranges from approximately 0.35 to 1.04 Å. Based on this, we estimated the $C_H$ range for Ni/Cu-N-C systems to be approximately 15-32 μF $cm^{-2}$. After determining this range, the results of the uncertainty test for $C_H$ are shown in **Fig. S12**. It can be observed that $C_H$ primarily influences the position of the activity volcano plots under different pH conditions by affecting the range of the electric field. Regardless of the parameter used, the reaction pathway involving bridge O* adsorption produces volcano plots that align more closely with experimental results compared to those involving atop O* adsorption. Furthermore, when $C_H$ = 25 μF $cm^{-2}$, the relative activity under acidic and alkaline conditions matches the experimental data most accurately.

To further provide evidence for bridge-site O adsorption on weak-binding M-N-C catalysts, the nitrogen K-edge XANES spectra of NiPc/CNT and CuPc/CNT catalysts are collected before and after

ORR test in a 0.1 M KOH electrolyte at 0.6 $V_{RHE}$, and the spectra are compared in **Fig. 6c** with that of the reference FePc/CNT catalyst, which is widely recognized to proceed ORR via the atop oxygen adsorption models. The deconvoluted spectra of the as-prepared catalysts all exhibit distinctive peaks that can be grouped between 398~406 eV (Peaks 1~4) and 406~414 eV (Peaks 5~7), which can be assigned to nitrogen $\pi^*$ and $\sigma^*$ resonance manifolds,[47] respectively. As expected, the spectrum of FePc/CNT catalyst remains largely unchanged after ORR test. The new feature emerged at *c.a.* 400-403 eV for NiPc/CNT and CuPc/CNT catalysts can be assigned to the N-O interaction[48]. Meanwhile, a N-O feature (~403.3 eV) can be located from the deconvoluted high-resolution N 1*s* XPS spectra of MPc/CNT (**Fig. 6d**) and M-COF366/CNT (**Fig. 6e**) catalysts (see detailed fitting parameters in **Table S5** in Supporting Information). Similarly, the reference FePc/CNT and Fe-COF366/CNT catalysts exhibit identical N 1*s* XPS spectra before and after ORR test. Moreover, these experimental results are in good agreement with the anti-bonding state in the bridging N-O configuration, as confirmed by our additional Projected Density of States (PDOS) calculations and Crystal Orbital Hamilton Population (COHP) analyses (**Fig. 6f-h**). In the experimental data shown in **Fig. 6b**, we observed that more electrons from nitrogen atoms in NiPc/CNT and CuPc/CNT shifted to the π* anti-bonding states after ORR testing. In the theoretical analysis, the Fe-N bonds in FePc show that all electrons in the N 2p orbital remain in a bonding state with Fe (**Fig. 6f**). In contrast, when N-O bonds form in CuPc or NiPc, the anti-bonding states (identified as π*) below the Fermi level become significantly pronounced. These findings provide compelling evidence supporting the accuracy of our new model and uncover a novel ORR mechanism for M-N-C catalysts with weakly adsorbed metal-N bonds.

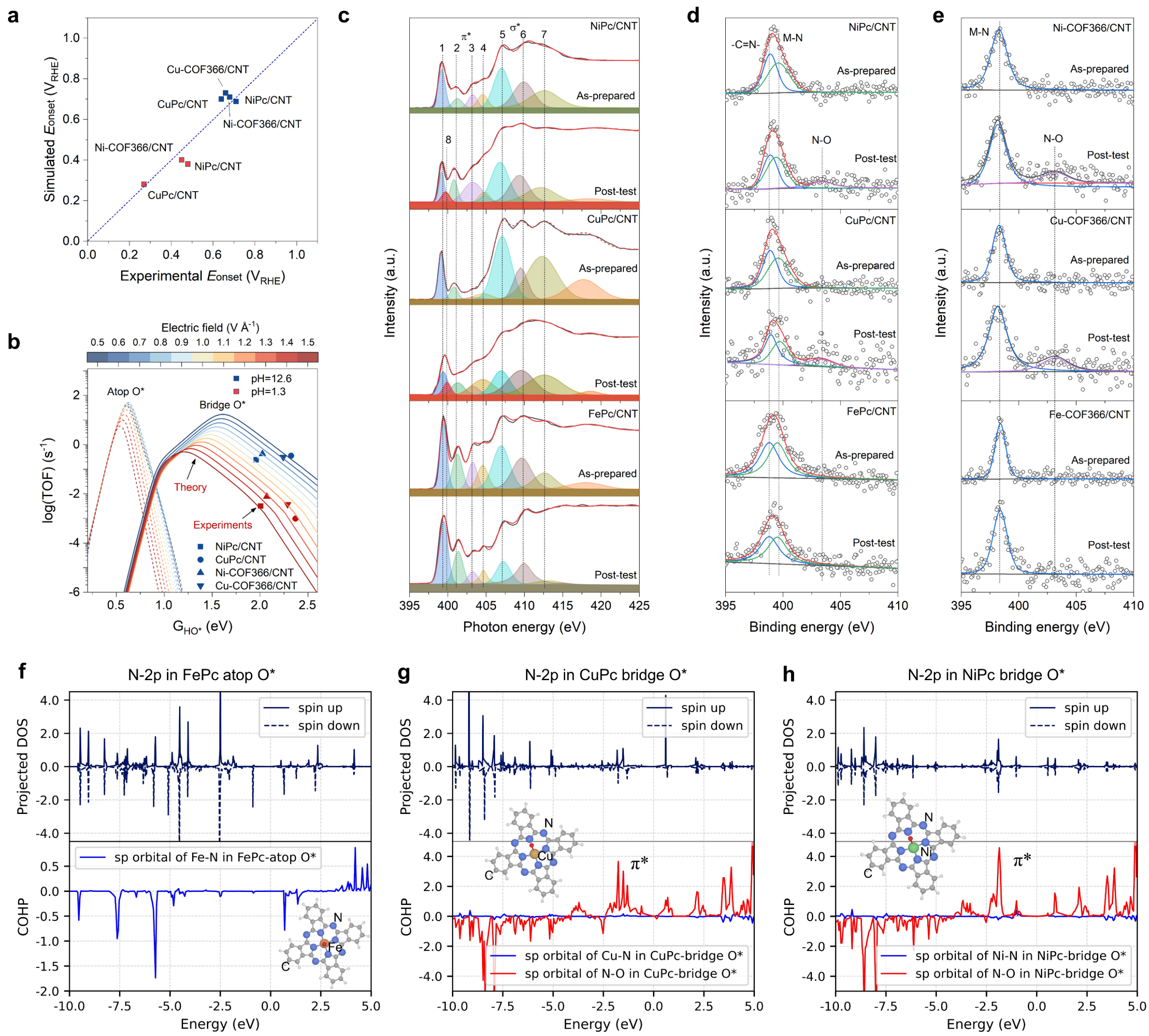

**Fig. 6. Experimental validation of the weak-binding M-N-C activity model.** (**a**) Onset potential correlation between the experiment (defined as the potential required to reach a $j_K$ of 0.1 mA cm$^{-2}$) and simulated results. (**b**) Comparison of the New Bridge-O* Model and the Typical Atop-O* Model in Matching Experimental TOFs. (**c-e**) X-ray spectroscopic analysis of catalysts before and after ORR test in a 0.1 M KOH electrolyte at 0.6 $V_{RHE}$. FePc/CNT or Fe-COF366/CNT reference catalysts are also tested for comparison. (**c**) N K-edge XAS spectra of the MPc/CNT catalysts. N1*s* XPS spectra of (**d**) MPc/CNT and (**e**) M-CNF366/CNT catalysts. (**f-h**) PDOS of the N 2p orbitals for (**f**) FePc with atop-O* adsorption, (**g**) CuPc with bridge-O* adsorption, and (**h**) NiPc with bridge-O* adsorption and corresponding COHP analysis, with corresponding COHP analysis below for metal-N or N-O bonds.

## Conclusion

In summary, this study advances the understanding of weak-binding M-N-C single-atom catalysts (SACs) by integrating comprehensive pH-dependent microkinetic modeling with experimental validations. Our findings reveal that, unlike the conventional preference for metal atop sites, atomic oxygen (O*) adsorption at M-N bridge sites in weak-binding SACs such as Ni/Cu-N-C plays a pivotal role in enhancing ORR performance. This bridge-site adsorption leads to significant variations in adsorption scaling relations, electric field responsiveness, and solvation effects, distinguishing the reaction pathway from that of moderate-binding SACs like Fe/Co-N-C.

The pH-electric field coupled model underscores that changes in dipole moment orientation, particularly for O* adsorption, mitigate pH dependence and lower kinetic barriers for ORR, therefore boosting catalytic efficiency in both acidic and alkaline conditions. Experimental confirmation through synchrotron spectroscopy, kinetic analyses, and X-ray absorption studies aligns closely with the proposed theoretical model, highlighting the presence of N-O bonds and increased electron density in anti-bonding state.

These insights challenge existing paradigms and underscore the need for further exploration of bridge-site interactions in weak-binding catalysts. The development of two distinct pH-dependent volcano models—considering both atop and bridge-site adsorptions—demonstrates improved alignment with experimental observations, validating our proposed mechanism. This work not only refines the mechanistic understanding of weak-binding M-N-C catalysts but also provides strategic directions for the design of next-generation electrocatalysts, tailored to exhibit robust performance across varying pH environments.

## Supporting Information

The Supporting Information is available free of charge at https://pubs.acs.org/doi/xxxx. All supplementary computational methods, experimental methods, figures, and tables mentioned in the main text can be found in the Supporting Information (PDF). The structures utilized for deriving scaling relations, analyzing electric field response, and investigating solvation effects are accessible at https://github.com/M-N-Cs/Weak-binding_SACs. The experimental data from large-scale datamining and the computational structures are available in the Digital Catalysis Platform (*DigCat*): https://www.digcat.org/.

## Acknowledgments

This research was supported by National Natural Science Foundation of China (No. 22309109), JSPS KAKENHI (Nos. JP23K13703 and JP24K23068) and the Hirose Foundation. We acknowledge the Center for Computational Materials Science, Institute for Materials Research, Tohoku University for the use of MASAMUNE-IMR (No. 202312-SCKXX-0203 and 202312-SCKXX-0207) and the Institute for Solid State Physics (ISSP) at the University of Tokyo for the use of their supercomputers. Hao Li and Li Wei acknowledge the financial and technical support provided by the University of Sydney under the International SDG Collaboration Program, the Australian Centre for Microscopy & Microanalysis (ACMM), and the Sydney Informatics Hub (SIH), and also acknowledge the computational resources provided by the National Computational Infrastructure (NCI, NCMAS). Li Wei acknowledges the funding support provided by the Australian Research Council Future Fellowship (Grant No. ARC-FT210100218). Di Zhang acknowledges the support provided by the Shanghai Jiao Tong University Outstanding Doctoral Student Development Fund and Siyuan-1 cluster supported by the Center for High Performance Computing at Shanghai Jiao Tong University.

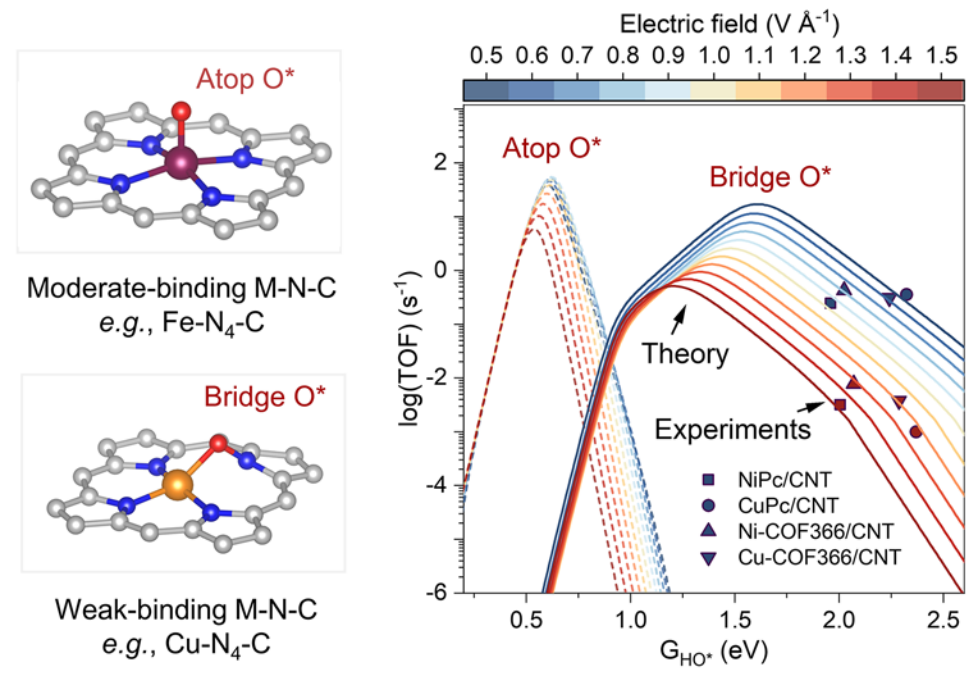

TOC Graphic